\documentclass[journal]{IEEEtran}
\usepackage{graphicx}

\usepackage{cite}

\ifCLASSINFOpdf
\else
\fi
\begin{document}
%
\title{Measurement of Charge Cloud Size\\ in X-ray SOI Pixel Sensors}
%
%
%

\author{Kouichi~Hagino,
        Kenji~Oono,
        Kousuke~Negishi,
        Keigo~Yarita,
        Takayoshi~Kohmura,\\
        Takeshi~G.~Tsuru,
        Takaaki~Tanaka,
        Hiroyuki~Uchida,
        Sodai~Harada,
        Tomoyuki~Okuno,
        Kazuho~Kayama,
        Yuki~Amano,
        Hideaki~Matsumura,
        Koji~Mori,
        Ayaki~Takeda,
        Yusuke~Nishioka,
        Kohei~Fukuda,
        Takahiro~Hida,
        Masataka~Yukumoto,
        Yasuo~Arai,
        Ikuo~Kurachi,
        Toshinobu~Miyoshi, and 
        Shunji~Kishimoto
\thanks{K.~Hagino,
        K.~Oono,
        K.~Negishi,
        K.~Yarita, and
        T.~Kohmura are with the Department of Physics, School of Science and Technology, Tokyo University of Science, Noda, Chiba 278-8510, Japan.}
\thanks{T.~G.~Tsuru,
        T.~Tanaka,
        H.~Uchida,
        S.~Harada,
        T.~Okuno,
        K.~Kayama, and
        Y.~Amano are with the Department of Physics, Faculty of Science, Kyoto University, Sakyo, Kyoto 606-8502, Japan.}
\thanks{H.~Matsumura is with the Kavli Institute for the Physics and Mathematics of the Universe (WPI), The University of Tokyo Institutes for Advanced Study, The University of Tokyo, 5-1-5 Kashiwanoha, Kashiwa, Chiba 277-8583, Japan}
\thanks{K.~Mori,
        A.~Takeda,
        Y.~Nishioka,
        K.~Fukuda,
        T.~Hida, and
        M.~Yukumoto are with the Department of Applied Physics, Faculty of Engineering, University of Miyazaki, Miyazaki, Miyazaki 889-2155, Japan.}
\thanks{ Y.~Arai, and
        T.~Miyoshi are with the Institute of Particle and Nuclear Studies (IPNS), High Energy Accelerator Research Organization (KEK), Tsukuba, Ibaraki 305-0801, Japan}
\thanks{I.~Kurachi is with the Department of Advanced Accelerator Technologies (AAT), High Energy Accelerator Research Organization (KEK), Tsukuba, Ibaraki 305-0801, Japan}
\thanks{S.~Kishimoto is with the Institute of Materials Structure Science (IMSS), High Energy Accelerator Research Organization (KEK), Tsukuba, Ibaraki 305-0801, Japan}
}

\maketitle

\begin{abstract}
We report on a measurement of the size of charge clouds produced by X-ray photons in X-ray SOI (Silicon-On-Insulator) pixel sensor named XRPIX. We carry out a beam scanning experiment of XRPIX using a monochromatic X-ray beam at 5.0~keV collimated to {\boldmath $\sim 10{\rm ~\mu m}$} with a 4-{\boldmath ${\rm \mu m\phi}$} pinhole, and obtain the spatial distribution of single-pixel events at a sub-pixel scale. The standard deviation of charge clouds of 5.0~keV X-ray is estimated to be {\boldmath $\sigma_{\rm cloud}=4.30\pm0.07{\rm ~\mu m}$}. Compared to the detector response simulation, the estimated charge cloud size is well explained by a combination of photoelectron range, thermal diffusion, and Coulomb repulsion. Moreover, by analyzing the fraction of multi-pixel events in various energies, we find that the energy dependence of the charge cloud size is also consistent with the simulation.
\end{abstract}


%
\IEEEpeerreviewmaketitle

%
%
%
%

 

\bstctlcite{IEEEexample:BSTcontrol}

\section{Introduction}

For all scientific applications of X-ray sensors, a correct understanding of the detector response is of great importance for the purpose of accurate interpretation of the experimental data. Particularly in pixel sensors, such as X-ray charge-coupled devices (CCDs), X-ray CMOS sensors, and so on, an accurate knowledge on charge sharing with adjacent pixels is essential because it has a large impact on the detector response. When an X-ray photon enters the pixel center and the pixel size is large enough, almost all the charge produced by the X-ray is collected by the pixel (single-pixel event). The energy and position of the incident X-ray photon is easily determined by the pulse height and the detected pixel position. On the other hand, when an X-ray photon enters near the pixel boundary, the charge is split into and shared with two or more pixels (multi-pixel event). In this case, the energy and position cannot be reconstructed without accurate knowledge of ``charge clouds'' produced by the incident X-ray. Thus, the size or shape of charge cloud produced by incident X-ray photons is one of the important parameters of detector response\cite{Miller2018}.

The impact of the charge cloud size on the detector response becomes larger in the sensors covering wide energy band. One of the main effects on the spectral response due to the charge spreading relates to the split threshold, which discriminates the split charge from the readout noise. Due to this split threshold, some fraction of charge-sharing events are not judged as multi-pixel event, leading to underestimation of the incident X-ray energy. This effect is relatively strong at lower energies because the ratio of split threshold to X-ray energy is large. In sensors using a thick depletion layer, necessary to extend the sensitivity to energies up to some tens of keV, the charge clouds spread more widely and the probability of charge sharing becomes larger. Thus, the charge cloud size is the essential parameter in the detector response for sensors with wide-band coverage.

\begin{figure}[tbp]
\begin{center}
\includegraphics[width=\hsize]{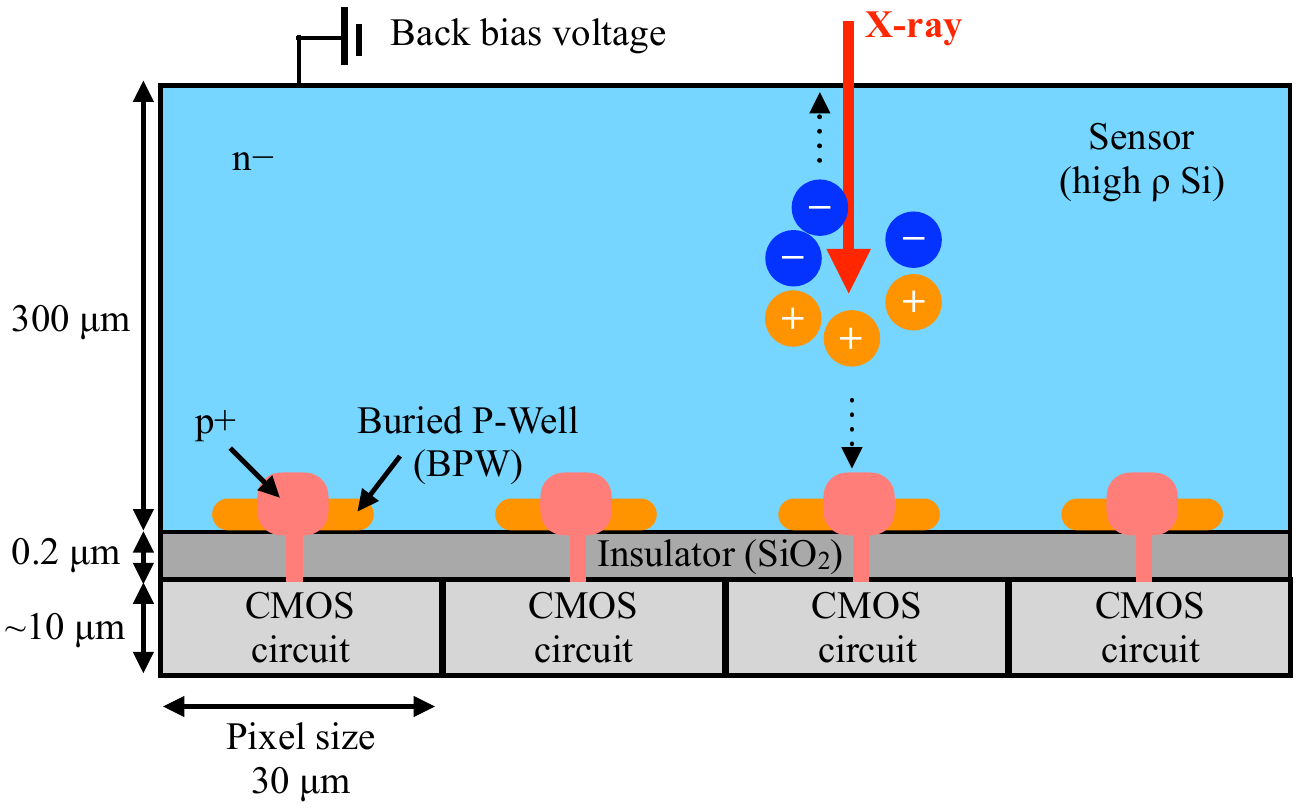}
\caption{Cross-sectional view of XRPIX. XRPIX is composed of high-resistivity sensor layer, SiO$_2$ insulator layer and CMOS circuit layer.
}
\label{fig:xrpix}
\end{center}
\end{figure}

\begin{figure*}[tbp]
\begin{center}
\includegraphics[width=0.8\hsize]{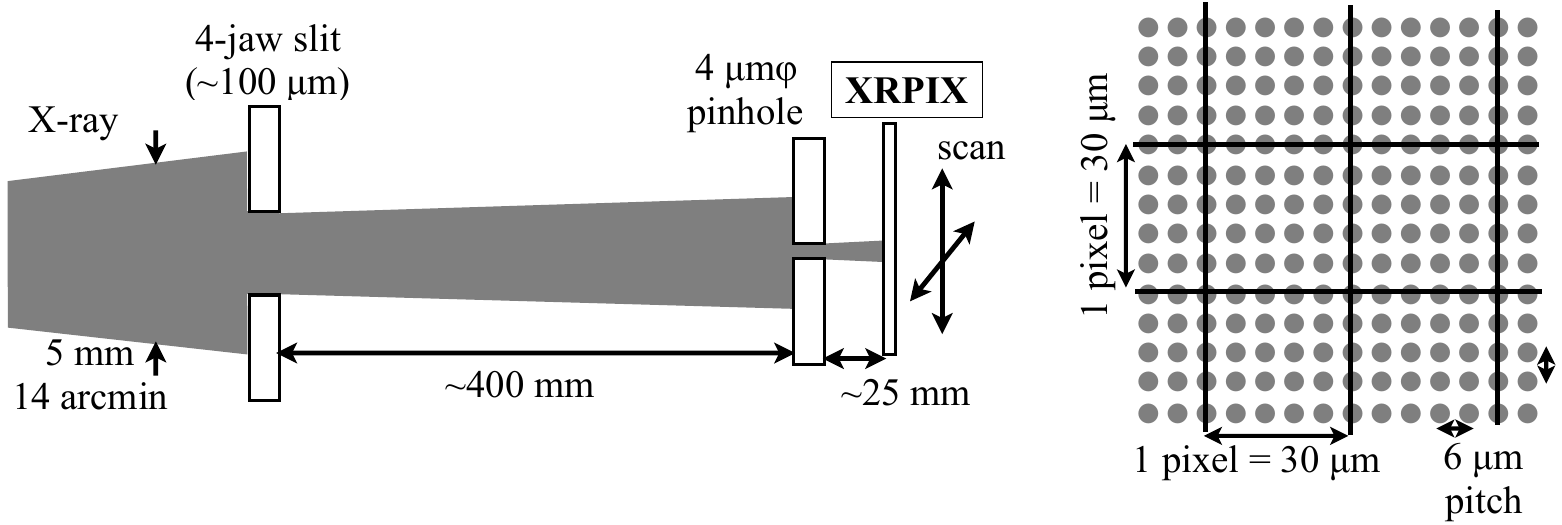}
\caption{Schematic picture of setup for the beam scanning experiment. X-ray beam was collimated with $4{\rm ~\mu m}\phi$ pinhole, and irradiated to XRPIX ({\it left}). XRPIX was scanned with $6{\rm ~\mu m}$ steps over $3\times3$ pixels ({\it right}).}
\label{fig:setup}
\end{center}
\end{figure*}

A future X-ray astronomical satellite ``FORCE'' aims at performing a wide-band imaging spectroscopy in 1--80~keV with a good angular resolution of $\sim10$~arcsecond in half power diameter\cite{Mori2016,Nakazawa2018}. The main detector of the FORCE satellite is a stacked detector composed of Si and CdTe layers. As the Si layer, we have been developing X-ray SOI (Silicon-On-Insulator) pixel sensors named XRPIX\cite{Arai2011,Tsuru2018}. For the FORCE satellite, depletion layer thicker than 200~$\mu$m, energy resolution better than 300~eV, and timing resolution better than 10~$\mu$s are required. Figure~\ref{fig:xrpix} shows a schematic picture of XRPIX. It is a monolithic active pixel sensor composed of high-resistivity Si sensor, thin SiO$_2$ insulator, and CMOS pixel circuits by utilizing the SOI technology. The high-resistivity Si sensor enables the thick depletion layer of a few hundreds of micrometers, which is essential to cover wide energy band. Since XRPIX is equipped with self-trigger capability in the pixel circuits, XRPIX can achieve a timing resolution better than $\sim10{\rm ~\mu s}$. 
Thanks to this high timing resolution, background can be drastically reduced by anti-coincidence technique with active shield counters\cite{Takahashi2006,Kokubun2006,Nakazawa2018a,Hagino2018}. Therefore, XRPIX has been considered to be a promising device for realizing future high-sensitivity X-ray observations.



Although measurements of the charge cloud size produced by X-ray photons were performed for X-ray CCDs\cite{Tsunemi1999}, such measurements have never been performed for XRPIX. On XRPIX, the detector response has been studied in Matsumura~et~al.~(2015)\cite{Matsumura2015} and Negishi~et~al.~(2018)\cite{Negishi2018}. These were focused on non-uniformity of the charge collection efficiency at a sub-pixel scale because it was one of the major issues in the development of XRPIX. So far, the charge cloud size in XRPIX has never been studied. Here, we report on a measurement of the charge cloud size in XRPIX. We also discuss their physical origin and the effect of non-uniform detection efficiency by comparing the experimental results with the detector response simulations.

\section{Beam Scanning Experiment}\label{sec:setup}
\begin{table}[t]
\caption{Specifications and Operational Parameters of XRPIX3b}
\begin{center}
\begin{tabular}{lr}
\hline\hline
Parameters & Value\\
\hline
Thickness of sensor layer & $300{\rm ~\mu m}$\\
Resistivity of sensor layer & $2{\rm ~k\Omega~cm}$\\
Pixel size & $30{\rm ~\mu m}\times 30{\rm ~\mu m}$\\
\hline
Back bias voltage & $350{\rm ~V}$\\
Temperature & $-70^\circ{\rm C}$\\
\hline\hline
\end{tabular}
\end{center}
\label{tab:spec}
\end{table}
In order to measure the charge cloud size in XRPIX, we carried out an X-ray beam scanning experiment at the beamline BL11B\cite{BL11B} of the synchrotron radiation facility, the Photon Factory of the High Energy Accelerator Research Organization (KEK) in Japan. We used a monochromatic X-ray beam at 5.0~keV for this experiment. The X-ray beam was collimated by a $4{\rm ~\mu m}\phi$ pinhole located at 25~mm from XRPIX. The pinhole was made of gold with a thickness of $90{\rm ~\mu m}$. Since the incident X-ray beam has a divergence angle of $\sim14{\rm ~arcmin}$ ($4{\rm~mrad}$) in full width, the beam size at XRPIX is larger than the pinhole size. To reduce the beam divergence, a 4-jaw slit was placed at the upstream of the pinhole. By moving the sensor with an X-Z stage, XRPIX was scanned with the collimated X-ray beam with 6~{\rm $\mu$m} steps over 90~{\rm $\mu$m} to cover $3\times3$ pixels, where the pixel size is $30{\rm ~\mu m}\times 30{\rm ~\mu m}$. The repeatability of the X-Z stage is $\leq 0.5{\rm ~\mu m}$, which is negligible in our measurement. More details of the experimental setup are described by Negishi~et~al.~(2018)\cite{Negishi2018}.

\begin{figure}[tbp]
\begin{center}
\includegraphics[width=0.8\hsize]{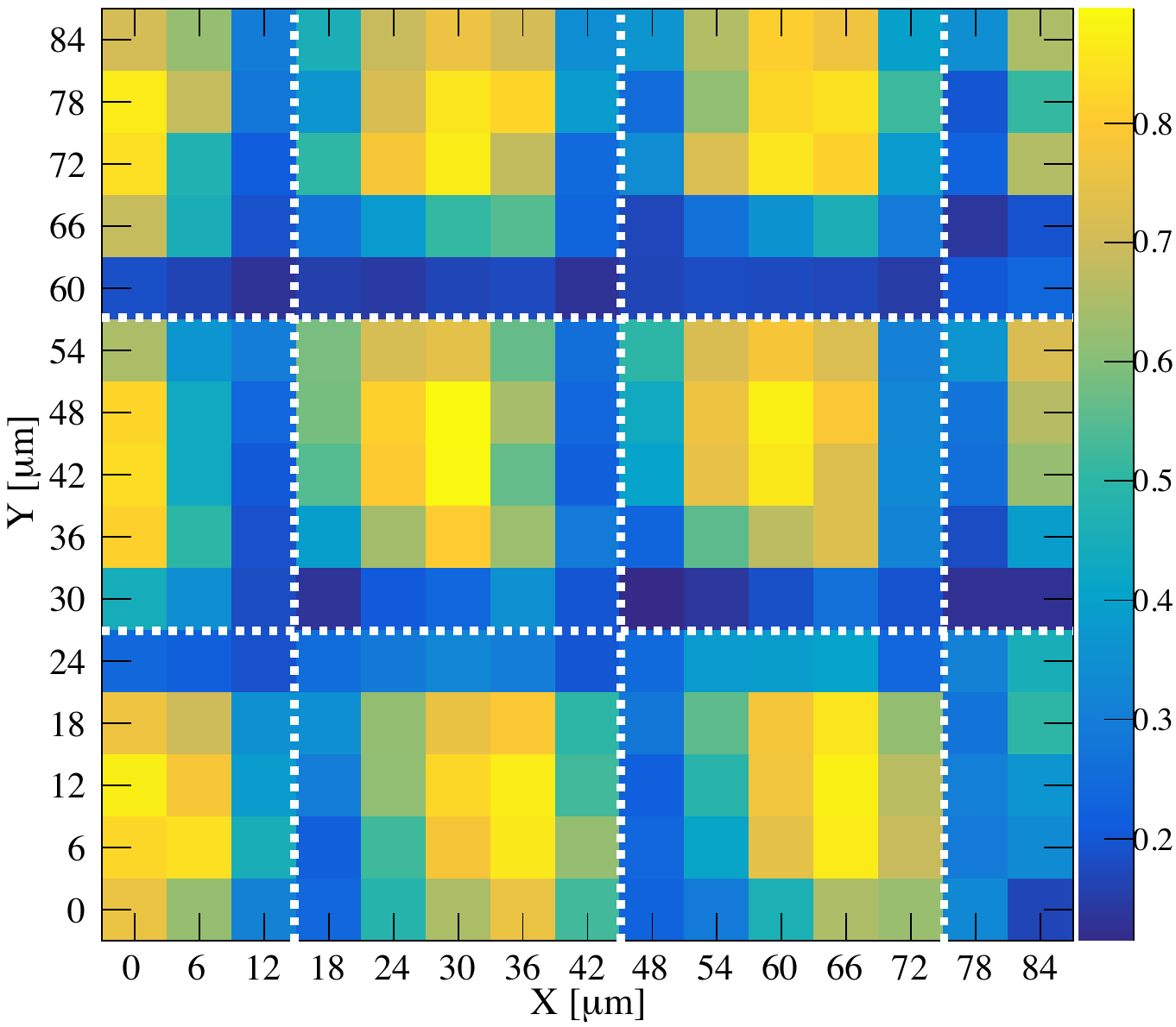}
\caption{Two-dimensional map of the fraction of the number of single-pixel events to those of the total X-ray events (color map) overlaid with estimated pixel boundaries (white dotted lines).}
\label{fig:map}
\end{center}
\end{figure}

In this experiment, we used XRPIX3b\cite{Takeda2015}, whose specifications and operational parameters in this experiment are listed in Table~\ref{tab:spec}. Its sensor layer is made of n-type Si with a resistivity of $\rho=2{\rm ~k\Omega~cm}$. The thickness of substrate in the sensor is $300{\rm ~\mu m}$.
Although a back-bias voltage of $\sim150{\rm ~V}$ is enough for full depletion, we applied a back-bias voltage of $350{\rm ~V}$ in order to maximize the charge collection efficiency. The sensor was cooled down to $-70^\circ{\rm C}$ to reduce the leakage current and thermal noise during the experiment. We irradiated the X-ray beam to the back side of XRPIX3b in order to avoid the total ionizing damage due to high intensity of the X-ray beam\cite{Yarita2018}. We did not use the trigger function, but periodically readout the entire sensor region like CCDs. It is because the spectroscopic performance is not so good in the event-driven readout mode using the trigger function due to cross-talks between circuit and sensor layer\cite{Ohmura2016}. We should note that such an issue has been resolved in the latest device\cite{Harada2018}.

The experimental data was analyzed with the method described in Ryu~et~al.~(2011)\cite{Ryu2011} and Nakashima~et~al.~(2012)\cite{Nakashima2012}. In the analysis, a pixel with a pulse height exceeding a predefined threshold is identified as an X-ray event. This threshold is referred to as the event threshold. For judgement of the charge sharing, we defined the other threshold called the split threshold. If the pulse height exceeds the split threshold in adjacent pixels, such events are classified as the multi-pixel events. On the other hand, all of the adjacent pixels have pulse heights below the split threshold, such events are classified as the single-pixel events.

We obtained a two-dimensional map of the single-pixel events as shown in Fig.~\ref{fig:map}. The color map represents fraction of the number of single-pixel event relative to those of both single- and multi-pixel event. Each position in the color map corresponds to each irradiation point in the scan. The fraction of single-pixel events becomes larger at the pixel center and becomes smaller at the pixel boundary. In Fig.~\ref{fig:map}, the single-pixel fraction shows a periodic dependence, and its period is roughly consistent with the pixel size. Therefore, we estimated pixel boundaries as white dotted lines in Fig.~\ref{fig:map}. This result demonstrates that the X-ray response of XRPIX3b was successfully measured at a sub-pixel scale.

\section{Estimation of Charge Cloud Size}
\subsection{Estimation with Single-pixel Fraction}\label{sec:single}
\begin{figure}[tbp]
\begin{center}
\includegraphics[width=\hsize]{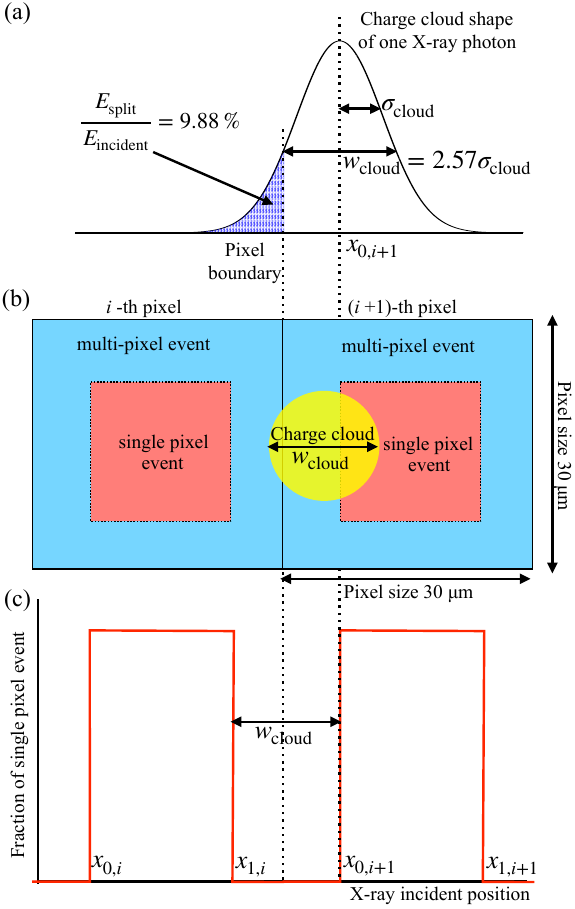}
\caption{Schematic picture of an interpretation of single-pixel event with a consideration of the charge cloud size. Single-pixel events are not detected at pixel boundary with a width of $w_{\rm cloud}$.}
\label{fig:method}
\end{center}
\end{figure}

We estimated the charge cloud size from the distribution of single-pixel events as shown in Fig.~\ref{fig:method}. We assume a Gaussian charge distribution with $\sigma_{\rm cloud}$ as shown in Fig.~\ref{fig:method}a. In this data analysis, single-pixel events are defined as the events in which charge above the split threshold $E_{\rm split}=0.494{\rm ~keV}$ is not detected in none of adjacent pixels. Since a ratio of the split threshold to the incident X-ray energy $E_{\rm incident}=5.0{\rm ~keV}$ is $E_{\rm split}/E_{\rm incident}=9.88\%$ in this experiment, the charge cloud size is regarded as $2.57\sigma_{\rm cloud}$ for the distribution of single-pixel fraction. Thus, we define this width as ``cloud width'' $w_{\rm cloud}=(2.57\pm 0.03)\sigma_{\rm cloud}$, where the uncertainty is propagated from that in energy calibration curve.


\begin{figure*}[tbp]
\begin{center}
\includegraphics[width=0.8\hsize]{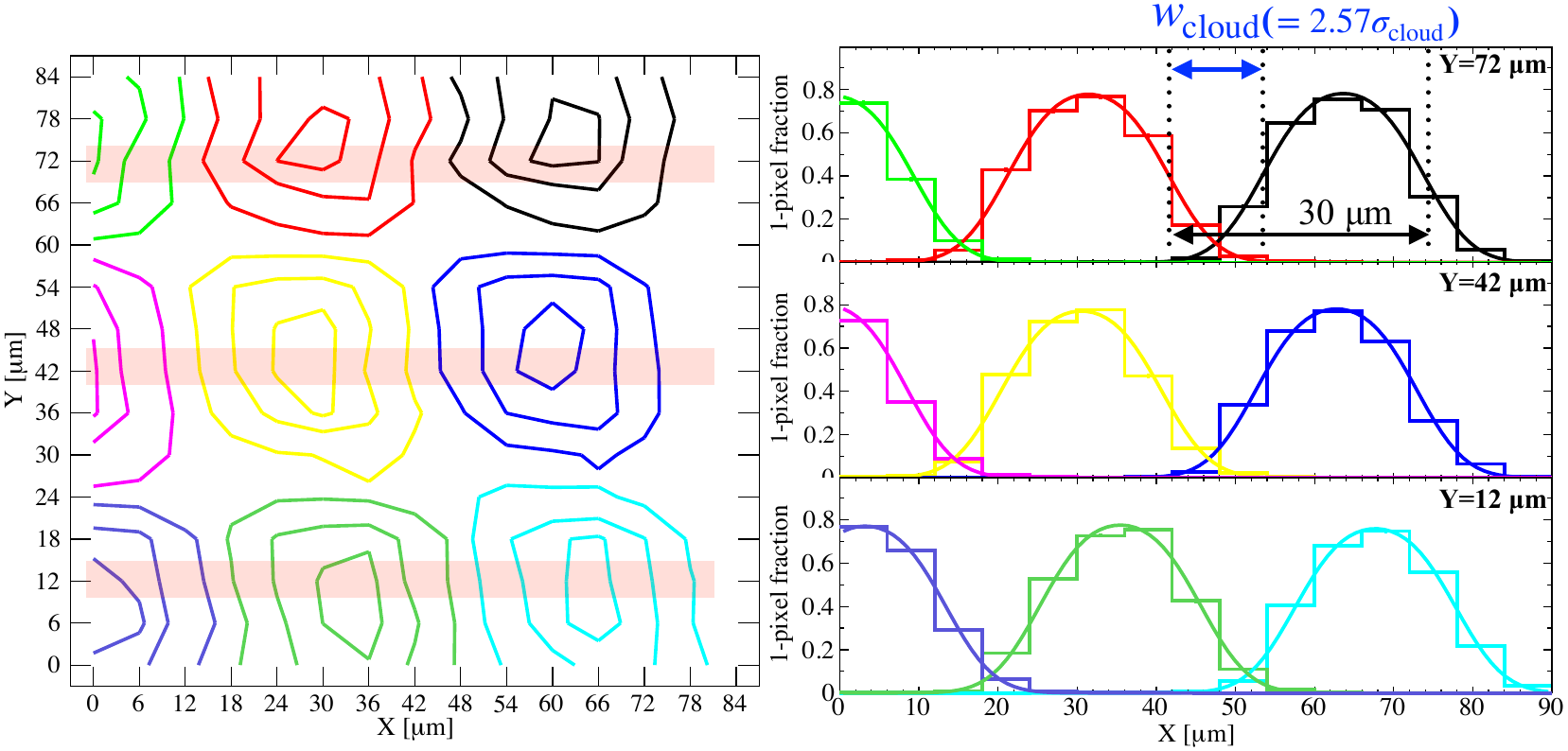}
\caption{{\it left}: Spatial distribution of single-pixel events for each pixel, obtained from the beam scanning experiment at the synchrotron radiation facility. The contour levels are the fractions of the number of single-pixel events at 0.2, 0.5 and 0.7. Each color represents individual pixel. Shaded regions represents the regions from which the slices shown in the right panel were extracted. {\it right}: Slices of the two-dimensional distribution of single-pixel events for each pixel. The best-fit model function is also plotted. Colors represent same pixels as those in the left panel.}
\label{fig:fitting}
\end{center}
\end{figure*}

Fig.~\ref{fig:method}b and \ref{fig:method}c schematically show the distribution of the single-pixel events as a function of the incident X-ray position. When the center of a charge cloud is separated from the pixel boundary by more than a half of the cloud width $w_{\rm cloud}$ (the red area in Fig.~\ref{fig:method}b), all the charge is collected into single pixel. On the other hand, when the charge is generated at pixel boundaries (the blue area in Fig.~\ref{fig:method}b), the charge cloud is split into and collected by multiple pixels. Thus, the spatial distribution of single-pixel events should be step functions with the cloud width $w_{\rm cloud}$ at the pixel boundary as shown in Fig.~\ref{fig:method}c. In experimental data, the step functions are smoothed with the finite beam size. By approximating the beam size with the Gaussian function, the one-dimensional distribution of single-pixel events for $i$-th pixel $f_{i}(x)$ becomes a superposition of error functions ${\rm erf}(x)$:
\begin{eqnarray}
f_{i}(x)&=&\frac{1}{2}\left[{\rm erf}\left(\frac{x-x_{0,i}}{\sqrt{2}\sigma_{\rm beam}}\right)+{\rm erf}\left(\frac{x_{1,i}-x}{\sqrt{2}\sigma_{\rm beam}}\right)\right],\label{eq:erf1}
\end{eqnarray}
where the distribution drops at $x=x_{0,i}, x_{1,i}$ with a slope controlled by the standard deviation of the beam profile $\sigma_{\rm beam}$. The edges of step function $x_{0,i}$ and $x_{1,i}$ are written as 
\begin{eqnarray}
x_{0,i}&=&x_0+iw_{\rm pixel}\\
x_{1,i}&=&x_0+(i+1)w_{\rm pixel}-w_{\rm cloud}.
\end{eqnarray}
Here, $x_{0}$ and $w_{\rm pixel}$ are the edge of the step function for $0$-th pixel ($x_{0,0}$) and the pixel size ($30{\rm ~\mu m}$), respectively.

As well as the beam size, the energy resolution smoothes the distribution of the single-pixel events. By considering the energy resolution with a typical noise level of $0.1{\rm ~keV}$ in full width half maximum, the ratio of the split threshold to the incident X-ray energy would have the uncertainty of $\Delta(E_{\rm split}/E_{\rm incident})=0.85\%$. It corresponds to the uncertainty of the edge position of the step function of $0.05\times\sigma_{\rm cloud}\simeq 0.2{\rm ~\mu m}$, which is much smaller than the beam size in this experiment. Thus, the contribution from the energy resolution is not included in Eq.~\ref{eq:erf1}.

We applied this method to the experimental data. As the first step, we separated the two-dimensional map of the fraction of single-pixel event (Fig.~\ref{fig:map}) into $3\times 3$ maps corresponding to individual pixels, utilizing the hit pixel address of each readout event. They are shown as contour maps in the left panel of Fig.~\ref{fig:fitting}, where the contour levels are 0.2, 0.5 and 0.7. Then, as the second step, we extracted one-dimensional distributions of single-pixel events for individual pixels from these $3\times 3$ maps as shown in the right panel of Fig.~\ref{fig:fitting}. These distributions are obtained by slicing the maps at three lines, which are shown as shaded regions overlaid on the contour map. Finally, all these distributions were simultaneously fitted with superposition of error functions. Free parameters in the fitting were normalizations, positions of lower edges of the leftmost pixels $x_0$, beam size $\sigma_{\rm beam}$, pixel size $w_{\rm pixel}$ and cloud width $w_{\rm cloud}$. The last three parameters ($\sigma_{\rm beam}$, $w_{\rm pixel}$, $w_{\rm cloud}$) were tied among all the distributions. The pixel size was used as a ruler to correct distortion in $w_{\rm cloud}$ and $\sigma_{\rm beam}$, which was probably due to the imperfect alignment of the beam axis and the sensor.

The fitting of the spatial distribution of single-pixel events provided the cloud width of $w_{\rm cloud}=11.04\pm 0.12{\rm ~\mu m}$. Using the relation between the cloud width $w_{\rm cloud}$ and the Gaussian standard deviation of charge clouds $\sigma_{\rm cloud}$, the standard deviation of charge clouds of $5.0{\rm ~keV}$ X-ray in XRPIX3b is estimated to be $\sigma_{\rm cloud}=4.30\pm0.07{\rm ~\mu m}$.
%

The fitting also provided an approximated profile of incident X-ray beam. The best-fit value of the standard deviation of the beam profile was $\sigma_{\rm beam}=4.49\pm0.05{\rm ~\mu m}$, which is significantly larger than the diameter of $4{\rm ~\mu m}$ of the pinhole. The value of $\sigma_{\rm beam}$ is affected by the repeatability of the X-Z stage, energy resolution, diffraction at the pinhole, and the beam divergence. As already described in this section and section~\ref{sec:setup}, contributions from the repeatability of the stage and the energy resolution are $\leq0.5{\rm ~\mu m}$ and $\simeq0.2{\rm ~\mu m}$, respectively. The diffraction of X-ray at $5.0{\rm ~keV}$ ($\lambda=0.25{\rm ~nm}$) is $\sim\arcsin\left(0.25{\rm ~nm}/4{\rm ~\mu m}\right)\simeq 0.2{\rm ~arcmin}$, enlarging the beam size on XRPIX by $\simeq1.5{\rm ~\mu m}$. The most dominant factor of the beam size is the beam divergence. Although we reduce it with the 4-jaw slit located at $400{\rm ~mm}$ from the pinhole (Fig.~\ref{fig:setup}), the precision of the slit size is $\sim100{\rm ~\mu m}$, which corresponds to an incident angle of $\sim1{\rm ~arcmin}$ at the pinhole. At XRPIX, these non-parallel X-rays enlarge the beam size to $\sim 10{\rm ~\mu m}$. Thus, the beam size of $\sigma_{\rm beam}=4.49\pm0.05{\rm ~\mu m}$ is reasonable value in our setup.

\subsection{Estimation with Mean Pulse Height}
We also estimated the charge cloud size from the distribution of mean pulse height. Basic idea of this method was originally proposed by Tsunemi~et~al.~(1999)\cite{Tsunemi1999}. The mean pulse height of each pixel is proportional to the amount of charge collected into the pixel. Thus, if we assume that the convolution of the incident X-ray beam shape and the charge cloud shape is a Gaussian function, the one-dimensional spatial distribution of the mean pulse height of $i$-th pixel is expressed as 
\begin{eqnarray}
f_{i}(x)&=&\int_{x_{0,i}}^{x_{1,i}}\frac{1}{\sqrt{2\pi}\sigma_{\rm conv.}}\exp\left[-\frac{(x-x')^2}{2{\sigma_{\rm conv.}}^2}\right]dx'\nonumber\\
&=&\frac{1}{2}\left[{\rm erf}\left(\frac{x-x_{0,i}}{\sqrt{2}\sigma_{\rm conv.}}\right)+{\rm erf}\left(\frac{x_{1,i}-x}{\sqrt{2}\sigma_{\rm conv.}}\right)\right],\label{eq:erf2}
\end{eqnarray}
where $x_{0,i}$ and $x_{1,i}$ are the pixel boundary of $i$-th pixel, and $\sigma_{\rm conv.}=\sqrt{{\sigma_{\rm cloud}}^2+{\sigma_{\rm beam}}^2}$ is the Gaussian standard deviation of convolution of the beam shape and the charge cloud shape. Similarly to Eq.~\ref{eq:erf1}, $x_{0,i}$ and $x_{1,i}$ depend on each other, whose relation is written as
\begin{eqnarray}
x_{0,i}&=&x_0+iw_{\rm pixel}\\
x_{1,i}&=&x_0+iw_{\rm pixel}+w_{\rm plateau},
\end{eqnarray}
where an additional parameter $w_{\rm plateau}$ is introduced. Since it represents the width between the edges of the error functions, it equals to the pixel size $w_{\rm pixel}$ in ideal conditions. We should note that the edge position ($x_{0,i}$, $x_{1,i}$) and the slope ($\sigma$) in Eq.~\ref{eq:erf2} represent different physical parameters from those in Eq.~\ref{eq:erf1}.

\begin{figure}[tbp]
\begin{center}
\includegraphics[width=0.8\hsize]{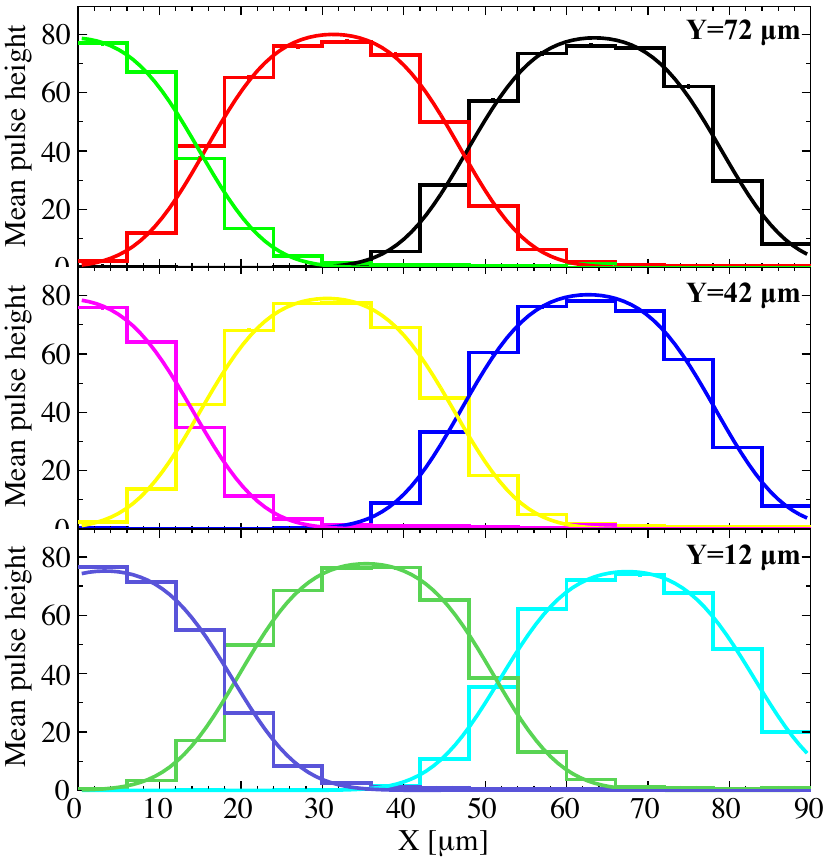}
\caption{One-dimensional spatial distribution of mean pulse height for each pixel. The best-fit model function is also plotted. Colors represent same pixels as those in Fig.~\ref{fig:fitting}. The slope of the error functions corresponds to root sum squared of the charge cloud size and the beam size.}
\label{fig:fitting_ph}
\end{center}
\end{figure}

Compared with the previous method using the single-pixel fraction, an advantage of this method is that the charge cloud size is independent from the split threshold. However, this method does not provide the charge cloud size directly, but convolution of the beam size and the charge cloud size. It requires an additional information of the incident X-ray beam size.

We applied this method to the experimental data as shown in Fig.~\ref{fig:fitting_ph}. Similarly to the fitting of the single-pixel fraction, three parameters $\sigma_{\rm conv.}$, $w_{\rm pixel}$ and $w_{\rm plateau}$ were tied among all the distributions shown in Fig.~\ref{fig:fitting_ph}, and $\sigma_{\rm conv.}$ and $w_{\rm plateau}$ were corrected using the pixel size. The best-fit value of the convolved standard deviation $\sigma_{\rm conv.}=\sqrt{{\sigma_{\rm cloud}}^2+{\sigma_{\rm beam}}^2}$ was $\sigma_{\rm conv.}=6.23\pm0.04{\rm ~\mu m}$. Using the beam size $\sigma_{\rm beam}=4.49\pm0.05{\rm ~\mu m}$ obtained in the fitting of the single-pixel fraction, we estimated the charge cloud size to be $\sigma_{\rm cloud}=\sqrt{{\sigma_{\rm conv.}}^2-{\sigma_{\rm beam}}^2}=4.32\pm0.08{\rm ~\mu m}$. It is consistent with the value obtained from the single-pixel fraction. The fitting also provided the plateau width of $w_{\rm plateau}=29.23\pm0.09{\rm ~\mu m}$, which is slightly smaller than the pixel size of $30{\rm ~\mu m}$. This small difference between the plateau size and the pixel size is due to the low detection efficiency at the pixel boundary, which will be discussed in section \ref{sec:nonuni}.

\section{Physical Model of the Detector Response}
\subsection{Model Description}
We compare the charge cloud size with physical model of the detector response in order to understand the physical origin of the charge cloud size. In the detector response model, we considered three physical mechanisms.The first one is photoelectron range. After the photoelectric absorption process, the photoelectron produced by the incident X-ray moves a finite length in the sensor until it loses all of their energy. Since the photoelectron range of 5.0~keV X-ray is $\sigma\simeq0.06{\rm ~\mu m}$\cite{Everhart1971}, its contribution on the charge cloud size is very limited in this energy. The second mechanism is thermal diffusion. The charge cloud spreads due to the thermal motion when it drifts to the sense nodes. The third mechanism is Coulomb repulsion, which is induced by the electric field created by the charge cloud itself. These effects are simulated by using a Monte-Carlo-based response generator ``ComptonSoft''\cite{Odaka2010}, in which the energy deposit in the sensor is simulated by utilizing the Geant4 toolkit\cite{Agostinelli2003,Allison2006}, and then the charge clouds are spread with a two-dimensional Gaussian depending on the distance to the readout nodes, and finally the pulse height of each electrode is calculated by randomizing the amount of collected charge with noise.

\begin{figure}[tbp]
\begin{center}
\includegraphics[width=1.1\hsize]{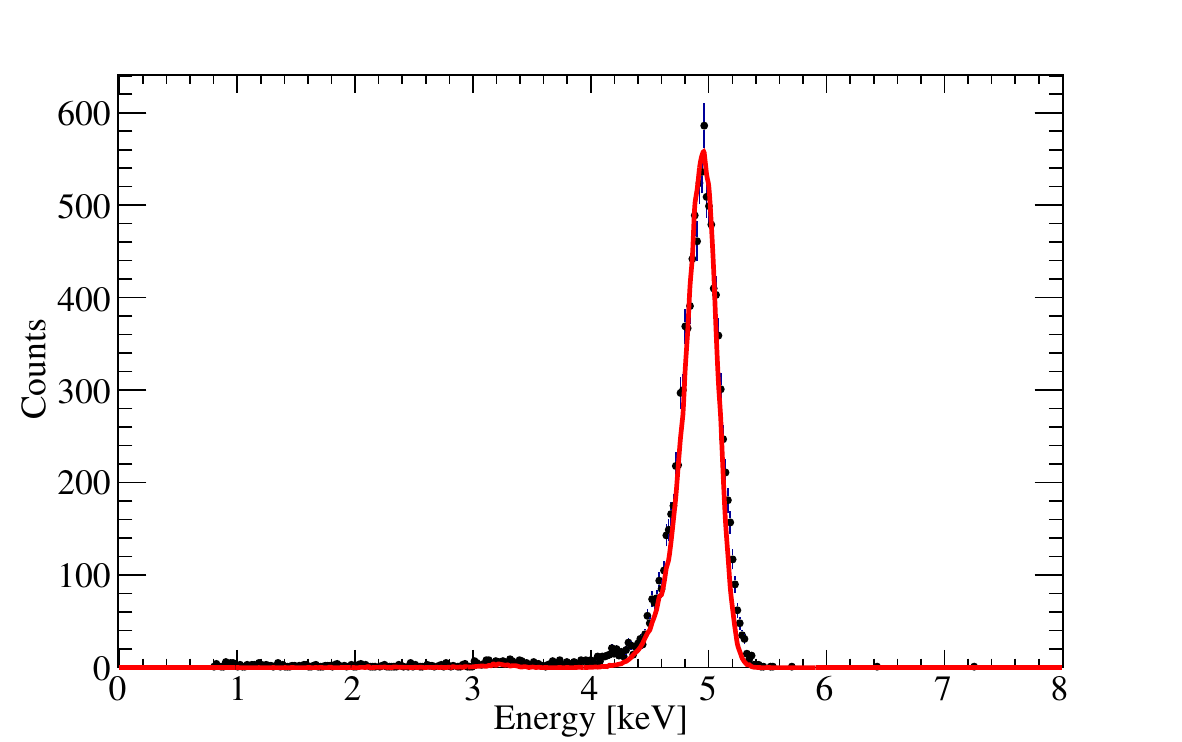}
\caption{Spectra of the simulations (red line) and the experimental data (black points) with an irradiation at the pixel center.}
\label{fig:spec}
\end{center}
\end{figure}

\begin{figure*}[tbp]
\begin{center}
\includegraphics[width=0.45\hsize]{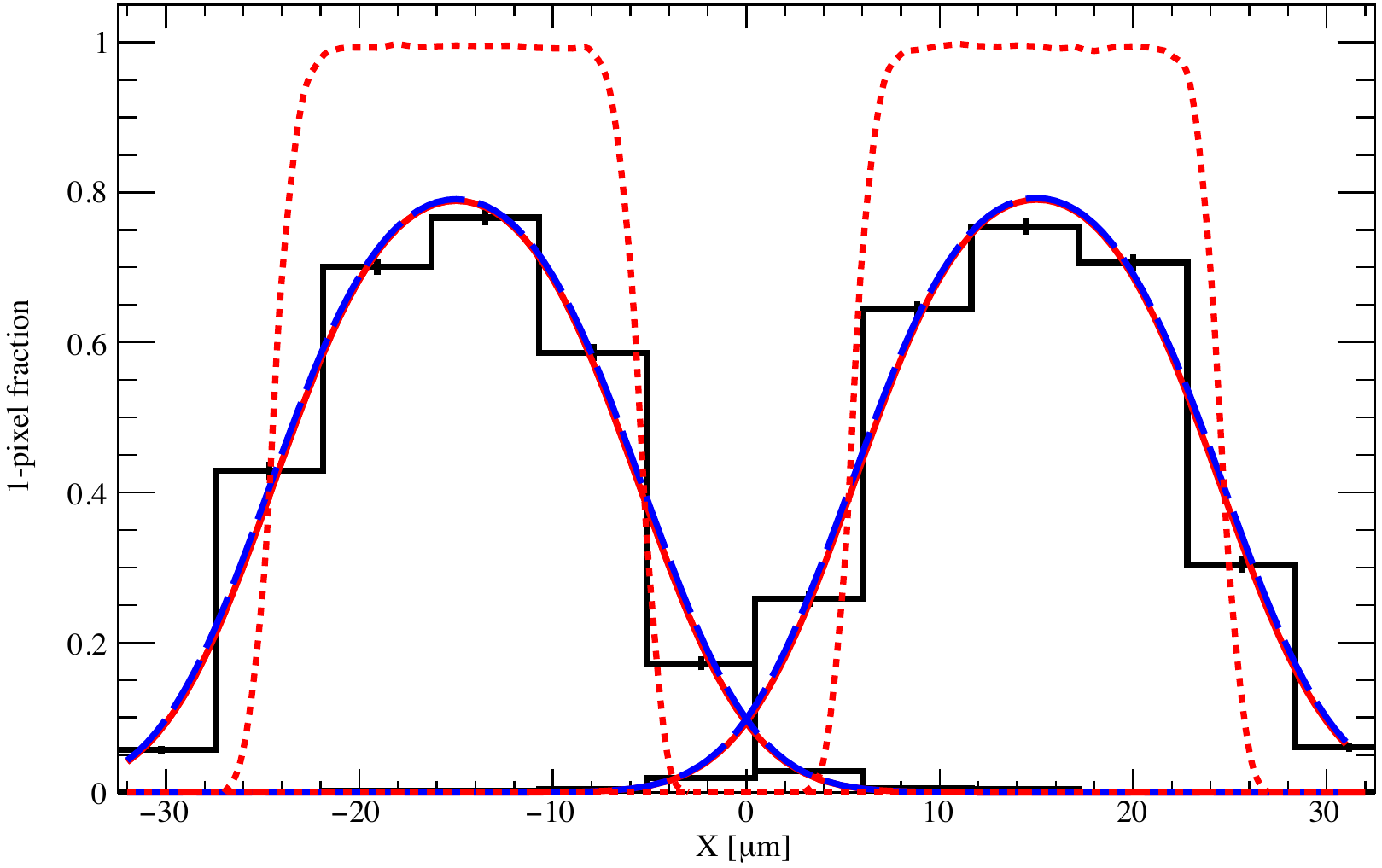}
\includegraphics[width=0.45\hsize]{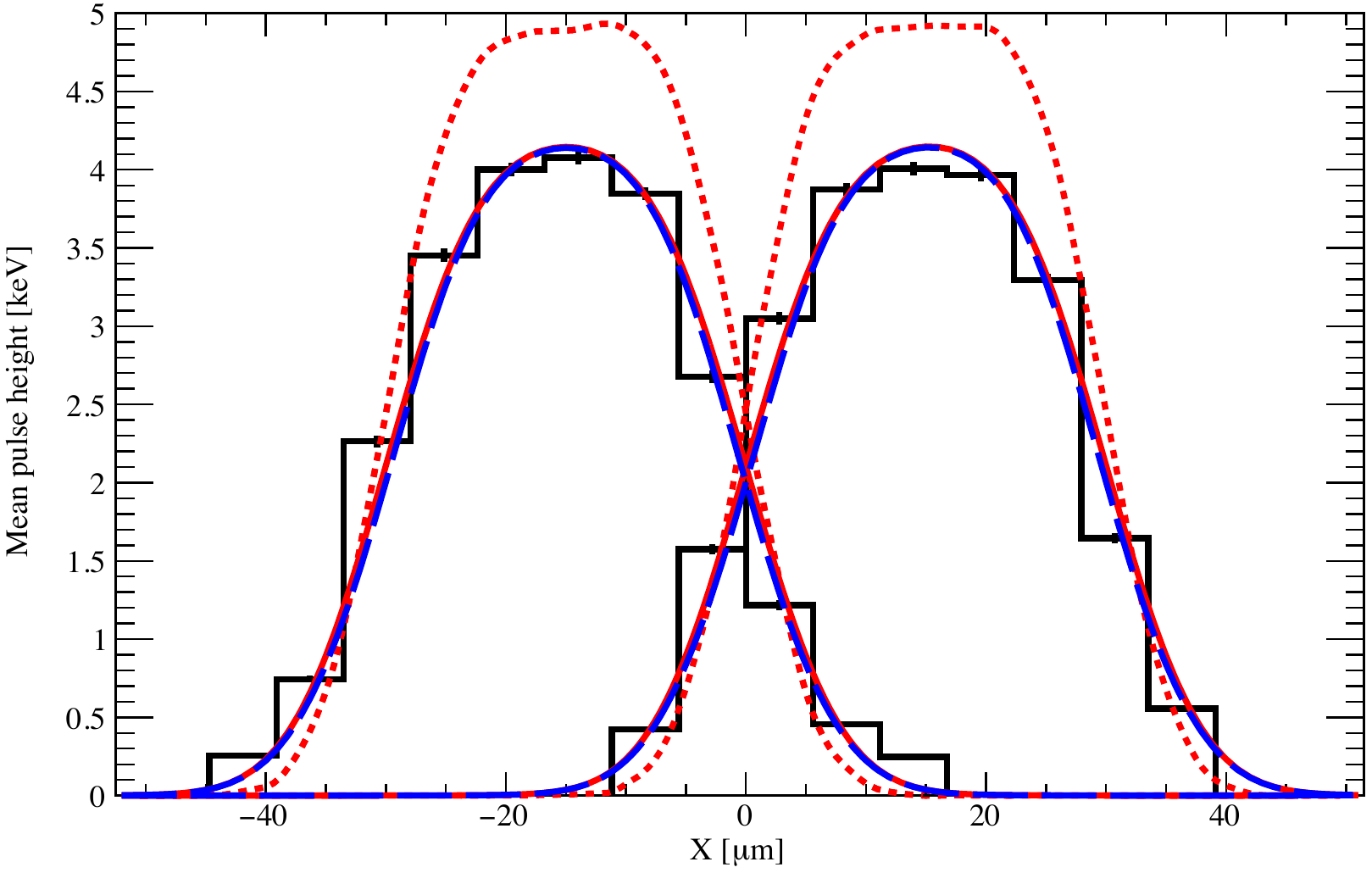}
\caption{One-dimensional spatial distributions of the single-pixel events (left) and the mean pulse height (right) obtained from the simulations (The red dotted curves). The red solid curves are the simulated distributions smoothed with the experimental beam profile. Those with consideration of the non-uniform detection efficiency are shown in the blue dashed curves. The distributions of the experimental data ($Y=72{\rm ~\mu m}$) were also shown as the black histograms. The position of the pixel boundary is defined as $x=0$ in this simulation.}
\label{fig:sim}
\end{center}
\end{figure*}

Since the effect of the photoelectron range was considered in the Geant4 Monte-Carlo simulation, we implemented charge spreading due to thermal diffusion and Coulomb repulsion. The width $\sigma$ of Gaussians of charge spreading after the drift time was calculated by solving a differential equation derived by Benoit \& Hamel~(2009)\cite{Benoit2009}, 
\begin{eqnarray}
\frac{\partial \sigma^2}{\partial t}=2D+\frac{\mu_{\rm p} Ne}{12\pi^{3/2}\varepsilon\sigma},\label{eq}
\end{eqnarray}
where $D=kT\mu_{\rm p}/e$ is the diffusion coefficient, $\mu_{\rm p}$ and $\varepsilon$ are the mobility of holes ($5.0\times10^2{\rm ~cm^2~V^{-1}~s^{-1}}$) and permittivity of Si ($1.04\times10^{-12}{\rm ~F~cm^{-1}}$), and $N$ is the number of charges generated by the X-ray photon.
In this calculation, the drift time $t_{\rm d}$ from carrier creation to the arrival to the sense nodes was calculated as
\begin{eqnarray}
t_{\rm d}&=&\int_{z_0}^0\frac{dz}{v_{\rm p}(z)},
\end{eqnarray}
where $z$ is the depth in the sensor layer measured from the readout nodes, $z_0$ is the position of charge creation, and $v_{\rm p}$ is the drift velocity of holes. We considered high-field effect of the drift velocity according to Caughey \& Thomas~(1967)\cite{Caughey1967},
\begin{eqnarray}
v_{\rm p}(z)&=&\frac{v_{\rm s}}{1+E_0/E(z)}.
\end{eqnarray}
Here, $v_{\rm s}=10^7{\rm ~cm~s^{-1}}$ is the saturation velocity, and $E_0=2\times10^{4}{\rm ~V~cm^{-1}}$ is a constant parameter. As an electric field in the sensor layer $E(z)$, we assumed a simple one-dimensional field expressed as
\begin{eqnarray}
E(z)=\frac{eN_{\rm d}}{\varepsilon}\left(z-\frac{d}{2}\right)-\frac{V}{d},
\end{eqnarray}
where $N_{\rm d}$ is the dopant density in the sensor layer ($2\times10^{12}{\rm ~cm^{-3}}$), $d$ is the thickness of the sensor layer ($300{\rm ~\mu m}$), and $V$ is the back bias voltage ($350{\rm ~V}$). This formula is a solution of the Poisson equation $\nabla^2\phi(z)=-eN_{\rm d}/\varepsilon$ under the boundary conditions of $\phi(0)=0$ and $\phi(d)=V$. The initial charge cloud size was assumed to be $\sigma_{\rm ini}=0.0171T_{\rm e}^{1.75}/2{\rm ~[\mu m]}$, where $T_{\rm e}{\rm ~[keV]}$ is the energy of the photoelectron\cite{Everhart1971}. Based on this calculation, the Coulomb repulsion increases the charge cloud size for 5.0~keV photons by a factor of 1.17 from that with only the thermal diffusion. Thus, we simply implemented the effect of the Coulomb repulsion by increasing the diffusion effect by this factor.

In order to accurately reproduce the experimental data with simulations, the simulated pulse height was randomized with noise. The noise was approximated with a Gaussian with $\sigma=\sqrt{{p_0}^2+p_1E}$, where the first term $p_0$ represents a readout noise and the second term $p_1E$ represents the Fano noise. We adopted $p_0=0.1{\rm ~keV}$ and $p_1=F\epsilon=0.36{\rm ~eV}$ on the assumption of a typical Fano factor of Si $F=0.1$\cite{Lepy2000} and the electron-hole pair creation energy $\epsilon=3.6{\rm ~eV}$. The resultant noise at 5.0~keV in FWHM is $\sim0.3{\rm ~keV}$, which is consistent with a typical energy resolution of XRPIX3b at 5.0~keV\cite{Takeda2015}. We also applied a similar analysis to those for the experimental data. Especially, we adopted the same values for the event and split thresholds as the experimental data analysis, in which $0.799{\rm ~keV}$ and $0.494{\rm ~keV}$ were used, respectively.

As an demonstration, the energy spectra of 5.0~keV X-ray from the simulation is shown in Fig.~\ref{fig:spec}. In the simulation, incident X-ray beam is irradiated at the pixel center with the beam profile obtained in section~\ref{sec:single}. Since the simulation is performed photon by photon under the consideration of charge spreading, the tail structure due to the charge sharing below the split threshold is reproduced. Thus, both of the energy resolution and the tail structure were well reproduced in the simulation.

\subsection{Comparison with the Experimental Data and Effect of Non-Uniform Detection Efficiency}\label{sec:nonuni}
The spatial distributions of the single-pixel events and the mean pulse height obtained by the simulations are shown in Fig.~\ref{fig:sim}. Since the incident X-ray is a pencil beam with an infinitesimal beam size in this simulation, the spatial distribution approaches in shape a step function. For the purpose of direct comparison with the experimental data, the simulated spatial distribution was smoothed with the experimental beam profile, which are shown in Fig.~\ref{fig:sim}. The simulated distribution closely matches the experimental distribution. The simulations provided the charge cloud size of $\sigma_{\rm cloud}=4.30{\rm ~\mu m}$ for $5.0{\rm ~keV}$ X-rays, which is consistent with the experimental values, $4.30\pm0.07{\rm ~\mu m}$ and $4.32\pm0.08{\rm ~\mu m}$. 

While the simulations assume full charge collection efficiency, it has been shown that the XRPIX3b actually experience some charge collection loss at the pixel boundaries. Such a charge loss issue was first reported by Matsumura~et~al.~(2014)\cite{Matsumura2014} in XRPIX1b. In this device, the X-ray pulse height at the pixel boundary was significantly lower than that at the pixel center. Since it was caused by electrical interference on the sensor layer from the pixel circuit, it was improved by rearrangement of the pixel circuit layout\cite{Matsumura2015}. Thanks to this improvement, there is almost no difference in the pulse heights between the pixel center and pixel boundary in XRPIX3b\cite{Negishi2018}. However, Negishi~et~al.~(2018)\cite{Negishi2018} found that there is still a little non-uniformity in the detection efficiency of XRPIX3b at a sub-pixel scale. Compared with the pixel center, the efficiencies are about $95\%$ at two-pixel boundary and about $75\%$ at four-pixel boundary.

\begin{figure}[tbp]
\begin{center}
\includegraphics[width=0.8\hsize]{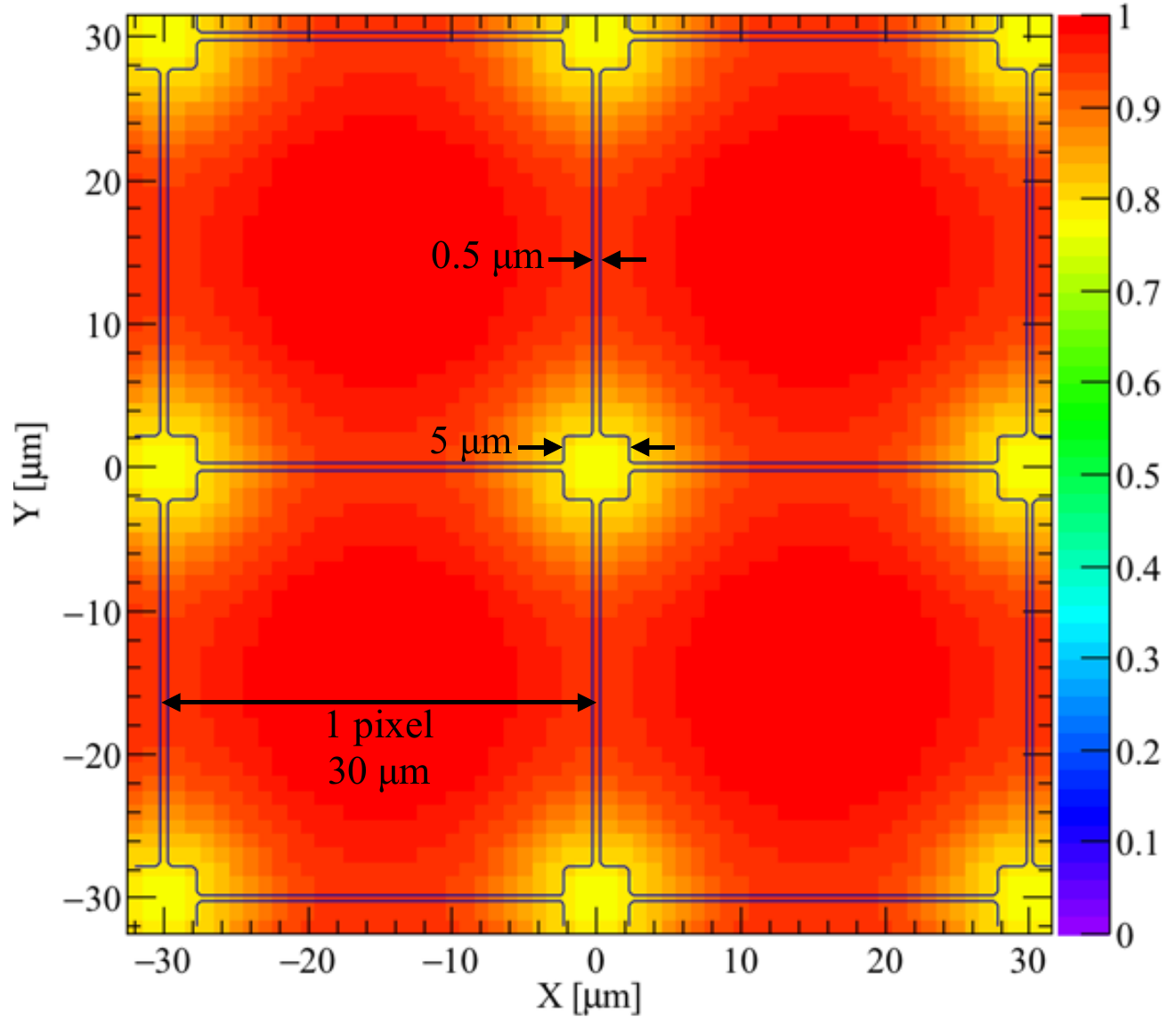}
\caption{Assumed configuration of the dead regions, where the incident X-rays are not detected. The regions surrounded by the black lines represent the dead regions. The color map shows the relative detection efficiency smoothed by the experimental beam profile (two-dimensional Gaussian with $\sigma_{\rm beam}=4.49{\rm ~\mu m}$), under the assumption of this dead region configuration.}
\label{fig:lossconfig}
\end{center}
\end{figure}

We evaluated the effect of the non-uniform detection efficiency on the measurement of the charge cloud size. We assumed line-like dead regions along two-pixel boundaries with a width of $0.5{\rm ~\mu m}$ and square dead regions with sides of $5{\rm ~\mu m}$ as shown in Fig.~\ref{fig:lossconfig}. This configuration is not physically evidenced, but is the simplest model reproducing the non-uniform detection efficiency seen in the experimental data. Smoothed with the two-dimensional Gaussian function with $\sigma_{\rm beam}=4.49{\rm ~\mu m}$, which is obtained from the experimental data, the relative efficiencies are $95.4\%$ at two-pixel boundary and $77.0\%$ at four-pixel boundary as shown in the color map in Fig.~\ref{fig:lossconfig}. These are consistent with the experimental results\cite{Negishi2018}. Considering this dead region configuration, we re-extracted the one-dimensional distributions of the single-pixel events and the mean pulse height, which are shown as the blue dashed lines in Fig.~\ref{fig:sim}. Since the two-pixel boundary dominated by the multi-pixel events is excluded as dead regions, the single-pixel fraction around the two-pixel boundary becomes slightly larger. It decreases the cloud width $w_{\rm cloud}$ by $2.6\%$. On the other hand, the standard deviation of the convolution of $\sigma_{\rm beam}$ and $\sigma_{\rm cloud}$ does not change in the distribution of the mean pulse height, while the plateau width $w_{\rm plateau}$ goes down to $29.20{\rm ~\mu m}$ because of the zero pulse height at the dead regions. Although the systematic uncertainty due to the non-uniform efficiency is larger than the statistical uncertainty, the experimental values matches those from the simulations with an accuracy of $\sim3\%$. Therefore, we conclude that the charge cloud size in XRPIX3b by the 5.0~keV photon is explained by photoelectron range, thermal diffusion, and Coulomb repulsion.

\section{Energy Dependence of Charge Cloud Size}
In addition to the scanned data obtained at the synchrotron radiation facility, we analyzed the data obtained with radioisotopes ($^{55}$Fe and $^{241}$Am) in our laboratory. Although X-rays were not collimated in this experiment, it is possible to investigate the energy dependence of the charge cloud size by comparing fractions of multi-pixel events of experimental data and simulations at each energy.

\begin{figure}[tbp]
\begin{center}
\includegraphics[width=\hsize]{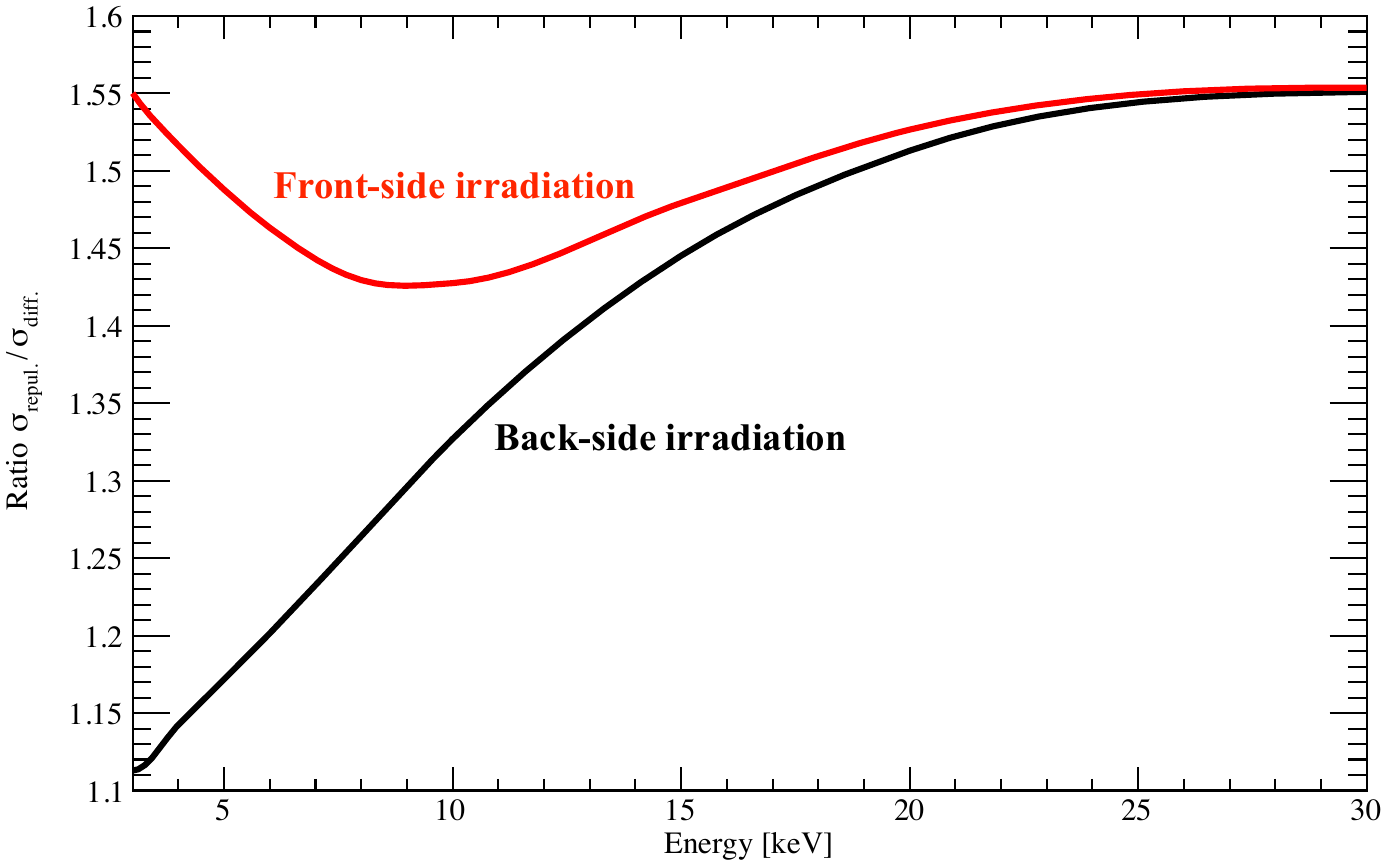}
\caption{Ratio of charge cloud sizes of thermal diffusion and Coulomb repulsion $\sigma_{\rm repul.}$ respect to thermal diffusion $\sigma_{\rm diff.}$. The ratio was calculated from Eq.~\ref{eq}. The ratio for back-side irradiation is shown in black and that for front-side irradiation is in red.}
\label{fig:repul}
\end{center}
\end{figure}

\begin{figure}[tbp]
\begin{center}
\includegraphics[width=\hsize]{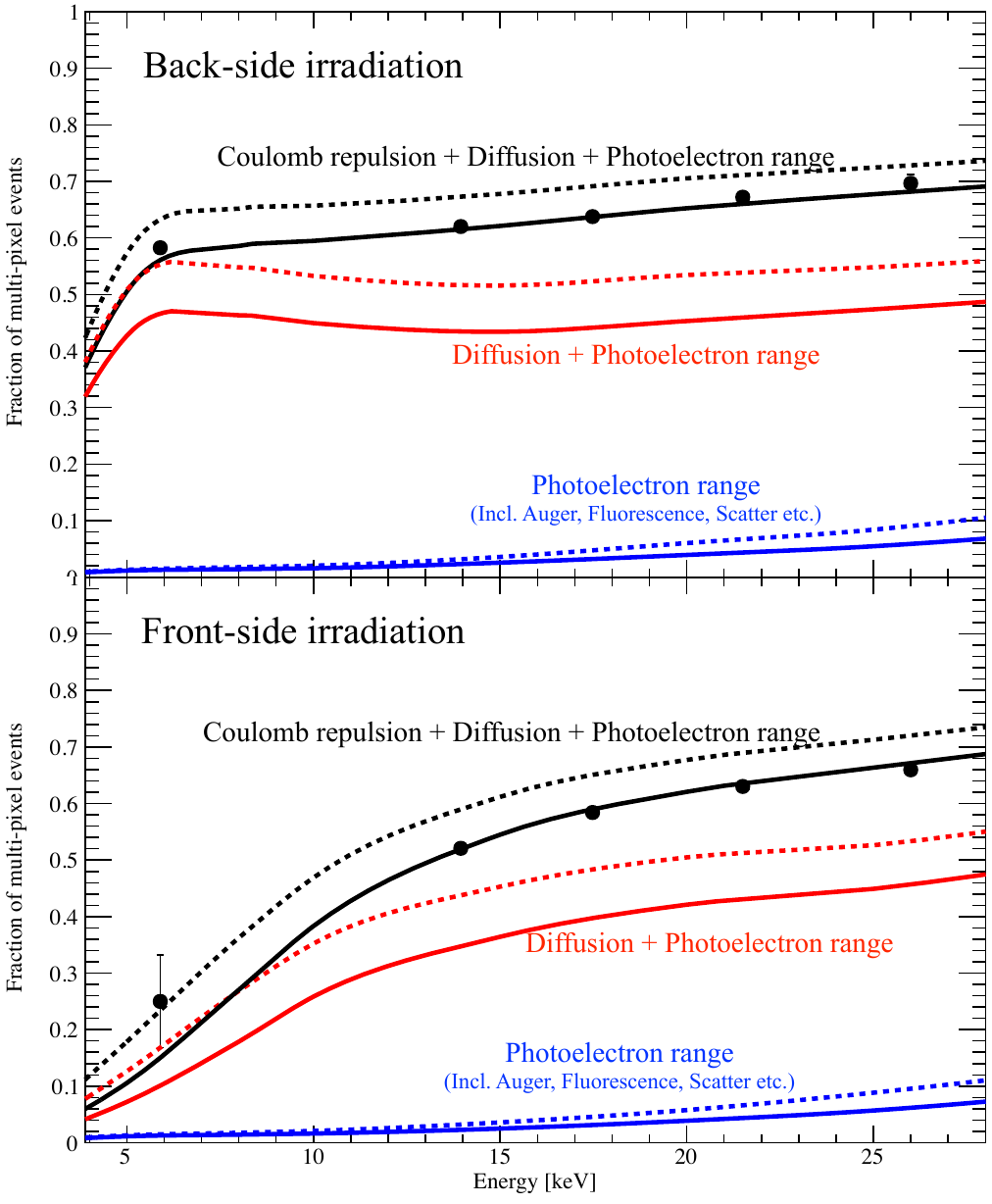}
\caption{Energy dependence of the multi-pixel fraction. The experimental data (black points) is consistent with the simulation (black solid line) considering photoelectron range, thermal diffusion and Coulomb repulsion with the assumption of the dead regions in Fig.~\ref{fig:lossconfig}. The solid lines indicate the simulated multi-pixel fractions with the dead region, while the dotted lines indicate those without the dead regions. The red and blue lines are from the simulations without Coulomb repulsion and thermal diffusion, respectively.}
\label{fig:ene}
\end{center}
\end{figure}

In order to implement the effect of Coulomb repulsion in various energies, we solved Eq.~\ref{eq} in various energies. Figure~\ref{fig:repul} shows the ratio of charge cloud sizes of thermal diffusion and Coulomb repulsion $\sigma_{\rm repul.}$ to thermal diffusion $\sigma_{\rm diff.}$. Based on Eq.~\ref{eq}, the charge spreading due to the Coulomb repulsion is roughly proportional to $({E_{\rm X}t_{\rm d}})^{1/3}$, while those of the thermal diffusion is proportional to ${t_{\rm d}}^{1/2}$. Here, $E_{\rm X}$ and $t_{\rm d}$ are the incident X-ray energy and the drift time, respectively. Thus, the ratio $\sigma_{\rm repul.}/\sigma_{\rm diff.}$ is roughly proportional to ${E_{\rm X}}^{1/3}{t_{\rm d}}^{-1/6}$. Since the electrostatic force is stronger for larger number of charges, the ratio increases as a function of X-ray energy. In case of front-side irradiation, for energies below 10~keV, the ratio becomes larger than at higher energies due to effect of the small drift time. As in the case of the simulation with 5.0~keV, we implemented the effect of Coulomb repulsion by increasing the charge spreading of thermal diffusion by these values.

We evaluated the fractions of multi-pixel events of four X-ray lines (Mn K$\alpha$ from $^{55}$Fe and Np L$\alpha$, L$\beta$, and L$\gamma$ from $^{241}$Am) and one gamma-ray line (26.3~keV from $^{241}$Am). Here, we used not only back-side irradiation data but also front-side irradiation data. By comparing the data with front-side and back-side irradiation, the drift time dependence of the charge cloud size as well as the energy dependence can be investigated. Fig.~\ref{fig:ene} shows the multi-pixel fractions of experimental data (black points) and the simulation results considering thermal diffusion, Coulomb repulsion and photoelectron range (black lines). As references, the multi-pixel fractions obtained through simulations not considering the Coulomb repulsion (red lines) and those with only photoelectron range (blue lines) are also shown. 

The event threshold was raised from that in the beam scanning experiment ($0.799{\rm ~keV}$) to $2.17{\rm ~keV}$ in this experiment. Since the number of pixels considered in the data analysis is much larger than the beam scanning experiment, the event threshold must be higher to avoid the miss-detection due to the read-out noise. On the other hand, the split threshold was unchanged because the split threshold is applied only for adjacent nine pixels in both of the experiments. Obviously, at energies below $4.34{\rm ~keV}$ ($2\times2.17{\rm ~keV}$), some fraction of two-pixel events must be undetected. Considering the energy resolution, the effect of the larger event threshold is seen at a little higher energies. In fact, the sharp drop below $\simeq6{\rm ~keV}$ in the back-side irradiation in Fig.~\ref{fig:ene} is due to the event threshold.

In this simulation, we also considered the non-uniform detection efficiency as shown in Fig.~\ref{fig:lossconfig}. The simulated multi-pixel fractions with this effect (solid lines) are smaller than those without this effect (dotted lines) by 10--20\%. This effect is mainly due to the dead regions at four-pixel boundaries. Compared with the simulations with the dead regions, the experimental data was successfully reproduced by considering the photoelectron range, thermal diffusion and Coulomb repulsion. Thus, the energy dependence and the drift time dependence of the charge cloud size in XRPIX3b were also well explained by these three physical mechanisms.

\begin{figure}[tbp]
\begin{center}
\includegraphics[width=1.1\hsize]{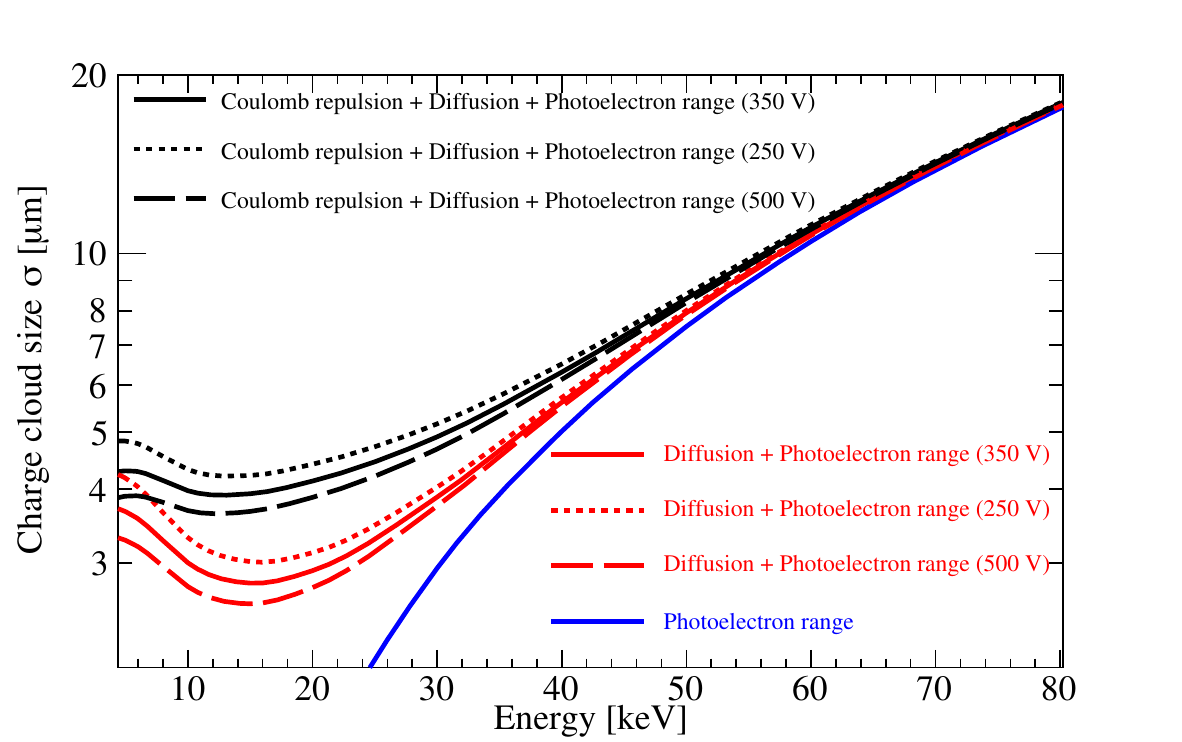}
\caption{Energy dependence of the charge cloud size $\sigma$ for different back bias voltages of $V_{\rm BB}=250, 350, 500 {\rm ~V}$. Back-side irradiation is assumed.}
\label{fig:cloudsize_ene}
\end{center}
\end{figure}

As a reference, the energy dependence of the charge cloud size up to 80~keV is shown in Fig.~\ref{fig:cloudsize_ene}. Contributions by photoelectron range and thermal diffusion are also overlaid. 
As already shown in Fig.~\ref{fig:repul} and Fig.~\ref{fig:ene}, the effects of thermal diffusion and Coulomb repulsion are dominant below 30~keV. On the other hand, at higher energies, photoelectron range plays a significant role because it rapidly increases in approximate proportion to ${E_{\rm X}}^{1.75}$\cite{Everhart1971}. In the figure, the charge cloud size under the different back bias voltages are also  shown. At lower energies, increase of the back bias voltage by a factor of two reduces the charge cloud size by a factor of $2^{1/3}$--$2^{1/2}$ because of the drift time dependence of thermal diffusion ($\propto {t_{\rm d}}^{1/2}$) Coulomb repulsion ($\propto {t_{\rm d}}^{1/3}$). Thus, at lower energies, the charge cloud size can easily change by the operational and device parameters, emphasizing the importance of the detailed modeling of the charge transportation and charge spreading.

\section{Conclusion}
We evaluated the X-ray response of XRPIX at a sub-pixel scale by irradiating XRPIX3b with 5.0~keV X-ray collimated with a 4~$\mu$m$\phi$ pinhole at KEK-PF. By utilizing the data obtained at KEK, the Gaussian standard deviation of charge clouds was estimated to be $\sigma_{\rm cloud}=4.30\pm0.07$~$\mu$m from one-dimensional distribution of the single-pixel events, and $4.32\pm0.08{\rm ~\mu m}$ from those of the mean pulse height. We constructed a detector response model considering photoelectron range, diffusion, and Coulomb repulsion, and obtained $\sigma_{\rm cloud}=4.30{\rm ~\mu m}$ from the simulations, which matches the experimental values with an accuracy of $\sim3\%$. The detector response model we developed also accurately reproduce the dependence of the multi-pixel events occurrence respect to the X-ray energy. It indicates that the energy dependence and the drift time dependence of the charge cloud size in XRPIX3b were able to be explained by these three physical mechanisms.
\section*{Acknowledgment}
We acknowledge the valuable advice and great work by the personnel of LAPIS Semiconductor Co., Ltd. This study was supported by MEXT/JSPS KAKENHI Grant-in-Aid for Scientific Research on Innovative Areas 25109002 (Y.A.), 25109003 (S.K.), 25109004 (T.G.T., T.T., K.M., A.T., and T.K.), Grant-in-Aid for Scientific Research (B) 25287042 (T.K.), Grant-in-Aid for Young Scientists (B) 15K17648 (A.T.), Grant-in-Aid for Challenging Exploratory Research 26610047 (T.G.T.) and Grant-in-Aid for JSPS Fellows 18J01417 (H.M.). This study was also supported by the VLSI Design and Education Center (VDEC), the University of Tokyo in collaboration with Cadence Design Systems, Inc., and Mentor Graphics, Inc.


\ifCLASSOPTIONcaptionsoff
  \newpage
\fi



\bibliographystyle{IEEEtran}
\bibliography{IEEEabrv,report}

\begin{thebibliography}{10}
\providecommand{\url}[1]{#1}
\csname url@samestyle\endcsname
\providecommand{\newblock}{\relax}
\providecommand{\bibinfo}[2]{#2}
\providecommand{\BIBentrySTDinterwordspacing}{\spaceskip=0pt\relax}
\providecommand{\BIBentryALTinterwordstretchfactor}{4}
\providecommand{\BIBentryALTinterwordspacing}{\spaceskip=\fontdimen2\font plus
\BIBentryALTinterwordstretchfactor\fontdimen3\font minus
  \fontdimen4\font\relax}
\providecommand{\BIBforeignlanguage}[2]{{%
\expandafter\ifx\csname l@#1\endcsname\relax
\typeout{** WARNING: IEEEtran.bst: No hyphenation pattern has been}%
\typeout{** loaded for the language `#1'. Using the pattern for}%
\typeout{** the default language instead.}%
\else
\language=\csname l@#1\endcsname
\fi
#2}}
\providecommand{\BIBdecl}{\relax}
\BIBdecl

\bibitem{Miller2018}
E.~Miller, R.~Foster, C.~Lage, G.~Prigozhin, M.~Bautz, C.~Grant, B.~Lamarr, and
  A.~Malonis, ``{The effects of charge diffusion on soft x-ray response for
  future high-resolution imagers},'' in \emph{Proceedings of SPIE}, vol. 10699,
  2018.

\bibitem{Mori2016}
K.~Mori, T.~G. Tsuru, K.~Nakazawa \emph{et~al.}, ``{A broadband x-ray imaging
  spectroscopy with high-angular resolution: the FORCE mission},'' in
  \emph{Proceedings of SPIE}, vol. 9905, 2016.

\bibitem{Nakazawa2018}
K.~Nakazawa, K.~Mori, T.~G. Tsuru \emph{et~al.}, ``{The FORCE mission: science
  aim and instrument parameter for broadband x-ray imaging spectroscopy with
  good angular resolution},'' in \emph{Proceedings of SPIE}, vol. 10699, 2018.

\bibitem{Arai2011}
Y.~Arai, T.~Miyoshi, Y.~Unno \emph{et~al.}, ``{Development of SOI pixel process
  technology},'' \emph{Nuclear Instruments and Methods in Physics Research
  Section A: Accelerators, Spectrometers, Detectors and Associated Equipment},
  vol. 636, no.~1, pp. S31--S36, 2011.

\bibitem{Tsuru2018}
T.~G. Tsuru, H.~Hayashi, K.~Tachibana \emph{et~al.}, ``{Kyoto's Event-Driven
  X-ray Astronomy SOI pixel sensor for the FORCE mission},'' in
  \emph{Proceedings of SPIE}, vol. 10709, 2018.

\bibitem{Takahashi2006}
T.~Takahashi, K.~Abe, M.~Endo \emph{et~al.}, ``{Hard X-Ray Detector (HXD) on
  Board Suzaku},'' \emph{Publications of the Astronomical Society of Japan},
  vol.~59, no. sp1, pp. S35--S51, 2007.

\bibitem{Kokubun2006}
M.~Kokubun, K.~Makishima, T.~Takahashi \emph{et~al.}, ``{In-Orbit Performance
  of the Hard X-Ray Detector on Board Suzaku},'' \emph{Publications of the
  Astronomical Society of Japan}, vol.~59, no. sp1, pp. S53--S76, 2007.

\bibitem{Nakazawa2018a}
K.~Nakazawa, G.~Sato, M.~Kokubun \emph{et~al.}, ``{Hard x-ray imager onboard
  Hitomi (ASTRO-H)},'' \emph{Journal of Astronomical Telescopes, Instruments,
  and Systems}, vol.~4, no.~02, pp. 021\,410--1, 2018.

\bibitem{Hagino2018}
K.~Hagino, K.~Nakazawa, G.~Sato \emph{et~al.}, ``{In-orbit performance and
  calibration of the Hard X-ray Imager onboard Hitomi (ASTRO-H)},''
  \emph{Journal of Astronomical Telescopes, Instruments, and Systems}, vol.~4,
  no.~02, pp. 021\,409--1, 2018.

\bibitem{Tsunemi1999}
H.~Tsunemi, J.~Hiraga, K.~Yoshita, and K.~Hayashida, ``{Measurement of the
  charge cloud shape produced by an X-ray photon inside the CCD using a mesh
  experiment},'' \emph{Nuclear Instruments and Methods in Physics Research
  Section A: Accelerators, Spectrometers, Detectors and Associated Equipment},
  vol. 421, no. 1-2, pp. 90--98, 1999.

\bibitem{Matsumura2015}
H.~Matsumura, T.~G. Tsuru, T.~Tanaka \emph{et~al.}, ``{Improving
  charge-collection efficiency of SOI pixel sensors for X-ray astronomy},''
  \emph{Nuclear Instruments and Methods in Physics Research Section A:
  Accelerators, Spectrometers, Detectors and Associated Equipment}, vol. 794,
  pp. 255--259, 2015.

\bibitem{Negishi2018}
K.~Negishi, T.~Kohmura, K.~Hagino \emph{et~al.}, ``{X-ray response evaluation
  in subpixel level for X-ray SOI pixel detectors},'' \emph{Nuclear Instruments
  and Methods in Physics Research Section A: Accelerators, Spectrometers,
  Detectors and Associated Equipment}, vol. 924, pp. 462--467, 2019.

\bibitem{BL11B}
``Bl-11b soft x-ray double-crystal monochromator station,''
  \url{http://pfwww.kek.jp/sxspec/sx/bl11b\_e.html}.

\bibitem{Takeda2015}
A.~Takeda, T.~Tsuru, T.~Tanaka \emph{et~al.}, ``{Improvement of spectroscopic
  performance using a charge-sensitive amplifier circuit for an X-ray
  astronomical SOI pixel detector},'' \emph{Journal of Instrumentation},
  vol.~10, no. C06005, 2015.

\bibitem{Yarita2018}
K.~Yarita, T.~Kohmura, K.~Hagino \emph{et~al.}, ``{Proton radiation damage
  experiment for X-ray SOI pixel detectors},'' \emph{Nuclear Instruments and
  Methods in Physics Research Section A: Accelerators, Spectrometers, Detectors
  and Associated Equipment}, vol. 924, pp. 457--461, 2019.

\bibitem{Ohmura2016}
S.~Ohmura, T.~G. Tsuru, T.~Tanaka \emph{et~al.}, ``{Reduction of cross-talks
  between circuit and sensor layer in the Kyoto's X-ray astronomy SOI pixel
  sensors with Double-SOI wafer},'' \emph{Nuclear Instruments and Methods in
  Physics Research Section A: Accelerators, Spectrometers, Detectors and
  Associated Equipment}, vol. 831, pp. 61--64, 2016.

\bibitem{Harada2018}
S.~Harada, T.~G. Tsuru, T.~Tanaka \emph{et~al.}, ``{Performance of the
  Silicon-On-Insulator pixel sensor for X-ray astronomy, XRPIX6E, equipped with
  pinned depleted diode structure},'' \emph{Nuclear Instruments and Methods in
  Physics Research Section A: Accelerators, Spectrometers, Detectors and
  Associated Equipment}, vol. 924, pp. 468--472, 2019.

\bibitem{Ryu2011}
S.~G. Ryu, T.~G. Tsuru, S.~Nakashima \emph{et~al.}, ``{First performance
  evaluation of an X-ray SOI pixel sensor for imaging spectroscopy and
  intra-pixel trigger},'' \emph{IEEE Transactions on Nuclear Science}, vol.~58,
  no. 5 PART 2, pp. 2528--2536, 2011.

\bibitem{Nakashima2012}
S.~Nakashima, S.~G. Ryu, T.~G. Tsuru \emph{et~al.}, ``{Progress in Development
  of Monolithic Active Pixel Detector for X-ray Astronomy with SOI CMOS
  Technology},'' \emph{Physics Procedia}, vol.~37, pp. 1373--1380, 2012.

\bibitem{Everhart1971}
T.~E. Everhart and P.~H. Hoff, ``{Determination of Kilovolt Electron Energy
  Dissipation vs Penetration Distance in Solid Materials},'' \emph{Journal of
  Applied Physics}, vol.~42, no.~13, pp. 5837--5846, 1971.

\bibitem{Odaka2010}
H.~Odaka, S.~Sugimoto, S.-n. Ishikawa \emph{et~al.}, ``{Development of an
  integrated response generator for Si/CdTe semiconductor Compton cameras},''
  \emph{Nuclear Instruments and Methods in Physics Research Section A:
  Accelerators, Spectrometers, Detectors and Associated Equipment}, vol. 624,
  no.~2, pp. 303--309, 2010.

\bibitem{Agostinelli2003}
S.~Agostinelli, J.~Allison, K.~Amako \emph{et~al.}, ``{Geant4―a simulation
  toolkit},'' \emph{Nuclear Instruments and Methods in Physics Research Section
  A: Accelerators, Spectrometers, Detectors and Associated Equipment}, vol.
  506, no.~3, pp. 250--303, 2003.

\bibitem{Allison2006}
J.~Allison, K.~Amako, J.~Apostolakis \emph{et~al.}, ``{Geant4 developments and
  applications},'' \emph{IEEE Transactions on Nuclear Science}, vol.~53, no.~1,
  pp. 270--278, 2006.

\bibitem{Benoit2009}
M.~Benoit and L.~Hamel, ``{Simulation of charge collection processes in
  semiconductor CdZnTe $\gamma$-ray detectors},'' \emph{Nuclear Instruments and
  Methods in Physics Research Section A: Accelerators, Spectrometers, Detectors
  and Associated Equipment}, vol. 606, no.~3, pp. 508--516, 2009.

\bibitem{Caughey1967}
D.~Caughey and R.~Thomas, ``{Carrier mobilities in silicon empirically related
  to doping and field},'' in \emph{Proceedings of the IEEE}, vol.~55, no.~12,
  1967.

\bibitem{Lepy2000}
M.~L{\'{e}}py, J.~Campbell, J.~Laborie, J.~Plagnard, P.~Stemmler, and
  W.~Teesdale, ``{Experimental study of the response of semiconductor detectors
  to low-energy photons},'' \emph{Nuclear Instruments and Methods in Physics
  Research Section A: Accelerators, Spectrometers, Detectors and Associated
  Equipment}, vol. 439, no. 2-3, pp. 239--246, 2000.

\bibitem{Matsumura2014}
H.~Matsumura, T.~G. Tsuru, T.~Tanaka, S.~Nakashima, S.~G. Ryu, A.~Takeda,
  Y.~Arai, and T.~Miyoshi, ``{Investigation of charge-collection efficiency of
  Kyoto's X-ray astronomical SOI pixel sensors, XRPIX},'' \emph{Nuclear
  Instruments and Methods in Physics Research Section A: Accelerators,
  Spectrometers, Detectors and Associated Equipment}, vol. 765, pp. 183--186,
  2014.

\end{thebibliography}
\end{document}